\pdfoutput=1 
\documentclass[a4paper,10pt]{article}
\usepackage{geometry}
\usepackage{graphics}  
\usepackage{graphicx}
\usepackage{subfigure}
\usepackage{amsmath}
\usepackage[amssymb]{SIunits}
\usepackage{nicefrac}
\usepackage[english]{babel}
\usepackage{lineno}
\usepackage{epstopdf}
\usepackage{stfloats}
\usepackage{verbatim}
\usepackage{url}
\usepackage{xcolor}
\usepackage{url}
\usepackage{setspace}

\newcommand{\Datura}{\ensuremath{\textrm{DATURA}}}

\newcommand{\Mimosa}{\ensuremath{\textrm{MIMOSA\,26}}}
\newcommand{\noise}{\ensuremath{\xi_{\textrm{n}}}}

\newcommand{\dz}{\ensuremath{\textrm{d}z}}

\newcommand{\eudet}{\ensuremath{\textrm{EUDET}}}

\newcommand{\eudaq}{\ensuremath{\textrm{EUDAQ}}}

\newcommand*{\notFOREPJ}{}%

\setpagewiselinenumbers
\modulolinenumbers[5]

\ifdefined\notFOREPJ
\else
\fi

\makeatletter
\renewcommand{\maketitle}{\bgroup\setlength{\parindent}{0pt}
\begin{flushleft}
  \vspace*{10mm}
  \textbf{\huge\sffamily\@title}
  \vspace{5mm}
   
  \large \@author
\end{flushleft}\egroup
}
\def\@xfootnote[#1]{%
  \protected@xdef\@thefnmark{#1}%
  \@footnotemark\@footnotetext}
\makeatother

\begin{document}
\ifdefined\notFOREPJ
\else
\fi







\title{Resolution studies with the DATURA beam telescope}
\author{
H.~Jansen${}^{\textrm{a,}}$\footnote[*]{Corresponding author: hendrik.jansen@desy.de}
\\
\vspace{3mm}
${}^{\textrm{a}}$ Deutsches Elektronen-Synchrotron DESY, Hamburg, Germany\\
\vspace{3mm}
}
\maketitle

\begin{abstract}
\noindent
Detailed studies of the resolution of a EUDET-type beam telescope are carried out using the $\Datura$ beam telescope as an example. 
The EUDET-type beam telescopes make use of CMOS $\Mimosa$ pixel detectors for particle tracking allowing for precise characterisation of particle sensing devices. 
A profound understanding of the performance of the beam telescope as a whole is obtained by a detailed characterisation of the sensors themselves. 
The differential intrinsic resolution as measured in a $\Mimosa$ sensor is extracted using an iterative pull method and various quantities that depend on the size of the cluster are discussed:
 the residual distribution, the intra-pixel residual-width distribution and the intra-pixel density distribution of track incident positions.\\


\end{abstract}



\section{Introduction}
\label{sec:intro}
\ifdefined\notFOREPJ
 
In the research and development of particle detectors, beam telescopes are indispensable instruments allowing for high precision characterisation of a device under test. 
The multitude of published results using beam telescopes reflect their importance. 
The integrated infrastructure initiative funded by the EU in the 6th framework programme 'structuring the European research area' included the EUDET project~\cite{EUDETwp}, 
 which realised the EUDET-type beam telescope, a high-resolution beam telescope using pixel detectors for test-beam studies~\cite{ref:eudetreport200902,JansenEPJ}. 
The, in total, seven installations of EUDET-type beam telescopes are deployed at different beam lines around the globe
 and rendered a multitude of detector studies possible, cf.\ references~\cite{Spannagel201671,AlipourTehrani:2133128,1748-0221-10-03-C03044,bib:IBLprototypes,bib:AFP3D2} among others.

In this work, a study of the intrinsic resolution as a function of the cluster size, i.e.\ the differential intrinsic resolution is presented
Making use of the few micrometre track resolution, the residual-width and density distributions as a function of the reconstructed incident position within a pixel cell is studied. 

\else
 
\fi
%

\section{Experimental set-up}
\label{sec:tscope}
\ifdefined\notFOREPJ
 
For this work, data have been taken at the \mbox{DESY-II} test-beam facilities~\cite{EUDET-2007-11}. 
The $\Datura$ beam telescope, described in detail in reference~\cite{JansenEPJ}, features six pixel detector planes equipped with fine-pitch $\Mimosa$ sensors~\cite{HuGuo2010480},
 the mechanics for precise positioning of the device under test and the telescope planes in the beam, a trigger logic unit (TLU) providing trigger capabilities and a data acquisition system.
Each $\Mimosa$ sensor consists of pixels sized $\unit{18.4}{\upmu\meter}\,\times\,\unit{18.4}{\upmu\meter}$, which are arranged in 1152 columns and 576 rows
 covering $\unit{21.2}{\milli\meter}\,\times\,\unit{10.6}{\milli\meter}$.
The average intrinsic resolution, i.e.\ the average over all cluster sizes, has been measured to be $\unit{3.24}{\upmu\meter}$ \cite{JansenEPJ}. 
The TLU is based on a commercial Spartan\,3 board~\cite{Spartan3}, and it features a coincidence unit with discriminator boards accepting up to four input signals, e.g.\ stemming from trigger scintillator devices. 
Providing a programmable logic, the TLU  takes a trigger decision based on its four input channels and issues asynchronous triggers subsequently to connected sub-systems. 

The chosen design meets most of the user requirements in terms of easy integration capabilities, spatial resolution and trigger rates. 
The telescope planes are designed and built to keep the amount of the material that the beam particles traverse as low as possible in order to achieve an excellent track resolution
 even at beam particle energies of a few GeV.

This article uses a right-handed coordinate system with the $z$-axis along the beam axis and the $y$-axis pointing downwards.
The origin of the coordinate systems is located at the centre of plane\,0. 
A narrow set-up with an equidistant plane spacing was chosen for this study, cf.~figure~\ref{fig:datura_sketch}, with $\dz = \unit{20}{\milli\meter}$. 
The sensor threshold per plane is tunable in multiples $\noise$ of their RMS noise. 
A sensor threshold setting of $\noise$ therefore implies a collected charge in a fired pixel of at least $\noise$ times the noise. 
 
\begin{figure}[tb]
	\center
	\ifdefined\notFOREPJ
	\includegraphics[trim= 80 170 230 110, width=.75\textwidth]{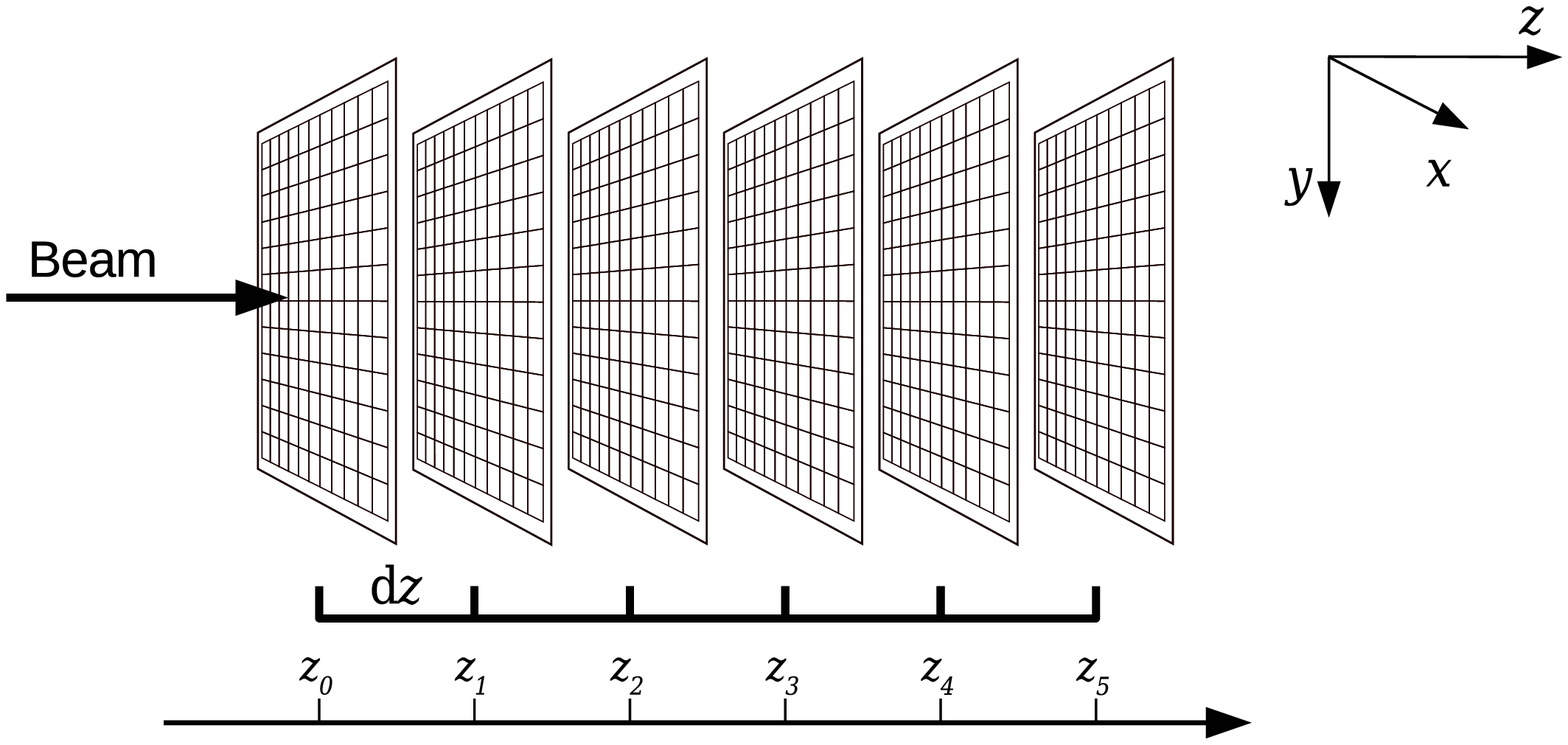}
	\else
	\includegraphics[trim= 80 130 230 60, width=.7\textwidth]{sketch_tscope.eps}
	\fi
	\caption[Sketch of the $\eudet$-type telescope set-up]{Sketch of the $\eudet$-type beam telescope with six $\Mimosa$ planes.}
	\label{fig:datura_sketch}
\end{figure}


\else
 
\fi

\section{Results \& discussion}
\label{sec:discussion}
\ifdefined\notFOREPJ
 
The results reported here are based on data obtained using the $\Datura$ beam telescope in the test-beam area~21 at DESY with a beam kinetic energy of 6\,GeV. 
The threshold of the $\Mimosa$ sensors were set to $\noise = 6$. 
The $\eudaq$ data acquisition framework and the EUTelescope analysis framework were used for data taking and offline data analysis, respectively. 
The clusters are formed by a sparse pixel clustering algorithm connecting all adjoining pixel hits.
Due to the lack of pulse height information from the binary $\Mimosa$ readout, a simple geometrical interpolation of the cluster centre is performed. 
Tracking was done using the concept of general broken lines~\cite{Blobel20111760,Kleinwort-2012}, which accounts for multiple scattering in the material traversed by the beam particles. 
Tracks have been formed in two ways, biased and unbiased, to be used for different purposes. 
Biased track fits use hits from all six pixel planes, i.e.\ also the hit information for the plane under investigation, whereas an unbiased track fit excludes the hit information for a given plane. 
More details about the analysis flow are presented in reference~\cite{JansenEPJ}.


\paragraph{Cluster size distribution}

Beam particles traversing the $\Mimosa$ sensors produce a variety of hit patterns therein. 
The measurement accuracy of the incident position depends on the size and the shape of the pattern. 
A hit with a cluster of size equal to one (CS1) is a hit with a single fired pixel in the cluster, a cluster size\,2 (CS2) hit contains information from two adjacent fired pixels,
 and accordingly for larger cluster sizes. 
In a CMOS sensor, free charges with low kinetic energy created close to or within the high-resistivity region are collected by the nearest or next-to-nearest read-out electrode. 
Therefore, with perpendicular incident, most of the produced cluster patterns are of size 1 to 4. 
A larger cluster size might e.g.\ arise from one or more additional noise pixels, by delta-electrons or a second beam particle passing in the vicinity of the first during a read-out cycle. 
The distribution of cluster size for a $\Mimosa$ sensor of the $\Datura$ beam telescope is shown in figure~\ref{fig:CSdist}. 
The distribution peaks for CS2 at about 28\,\%, the fraction of clusters larger than 4 decreases towards larger clusters. 
Clusters larger than six make up for less than 7\,\% of the the total distribution, the mean cluster size amounts to 3.28.
The mean for cluster sizes 1 to 4 yields 2.54.

\begin{figure}[tb]
	\center
	\ifdefined\notFOREPJ
	\includegraphics[width=.49\textwidth]{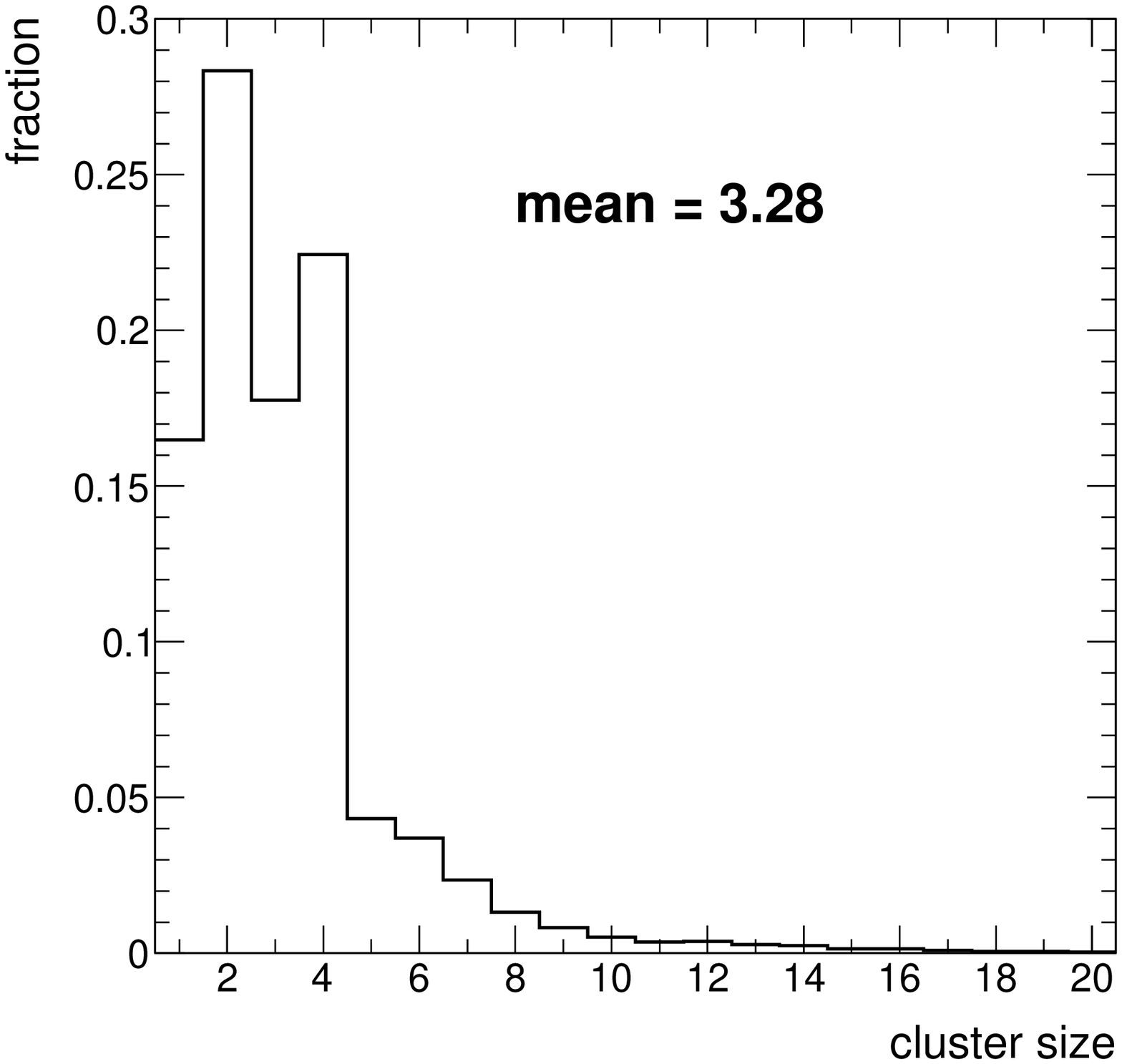}\put(-37,155){(a)}
	\includegraphics[width=.49\textwidth]{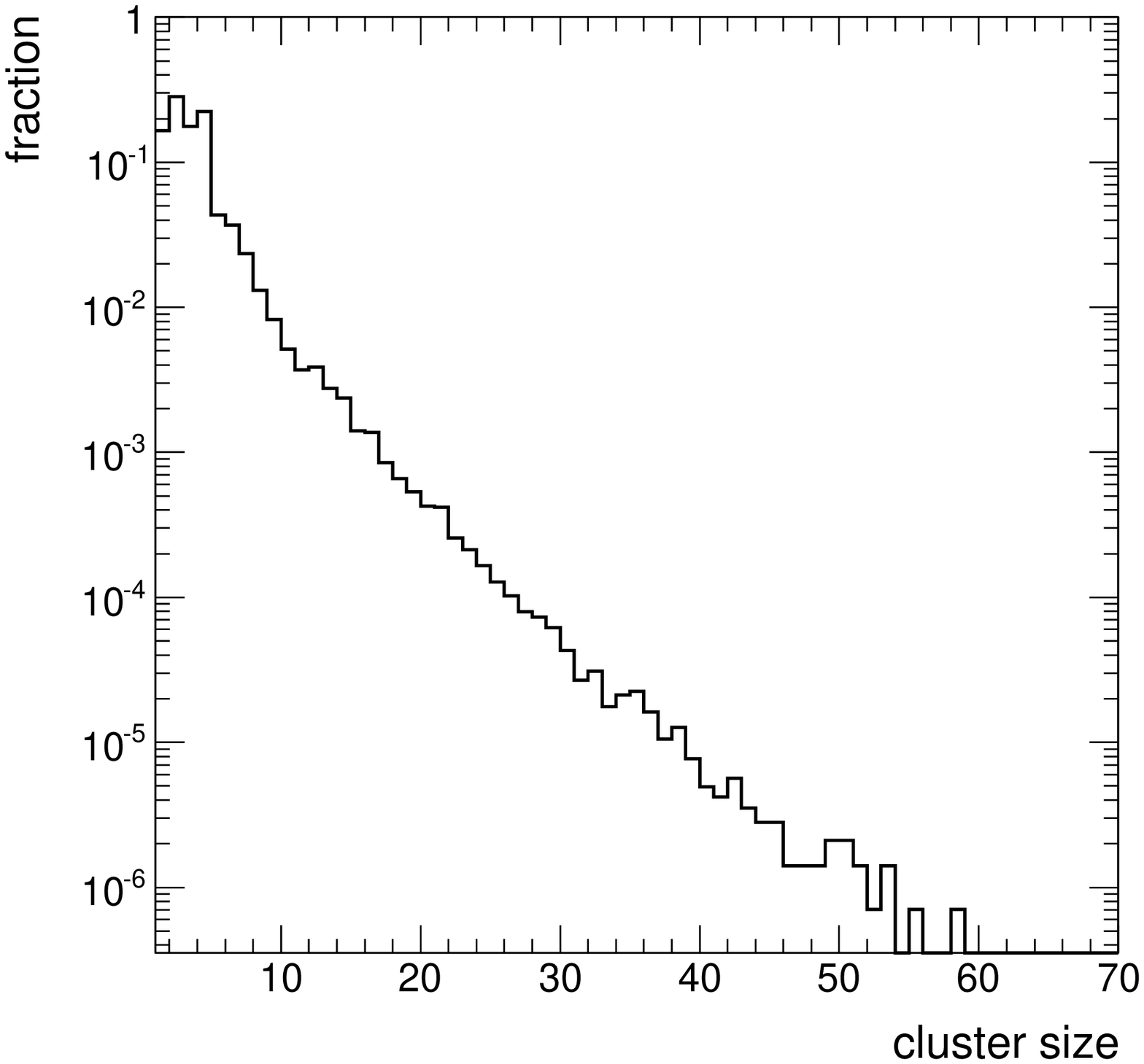}\put(-37,155){(b)}
	\else
	\includegraphics[width=.8\textwidth]{CSdist_plane3.eps}
	\fi
	\caption[Cluster size distribution]{The normalised cluster size distribution on (a) a linear and (b) a logarithmic scale is shown. 
	Error bars have been omitted as they are negligible for cluster sizes $<15$.}
	\label{fig:CSdist}
\end{figure}

\begin{figure}[b!]
	\center
	\ifdefined\notFOREPJ
	\includegraphics[width=.49\textwidth]{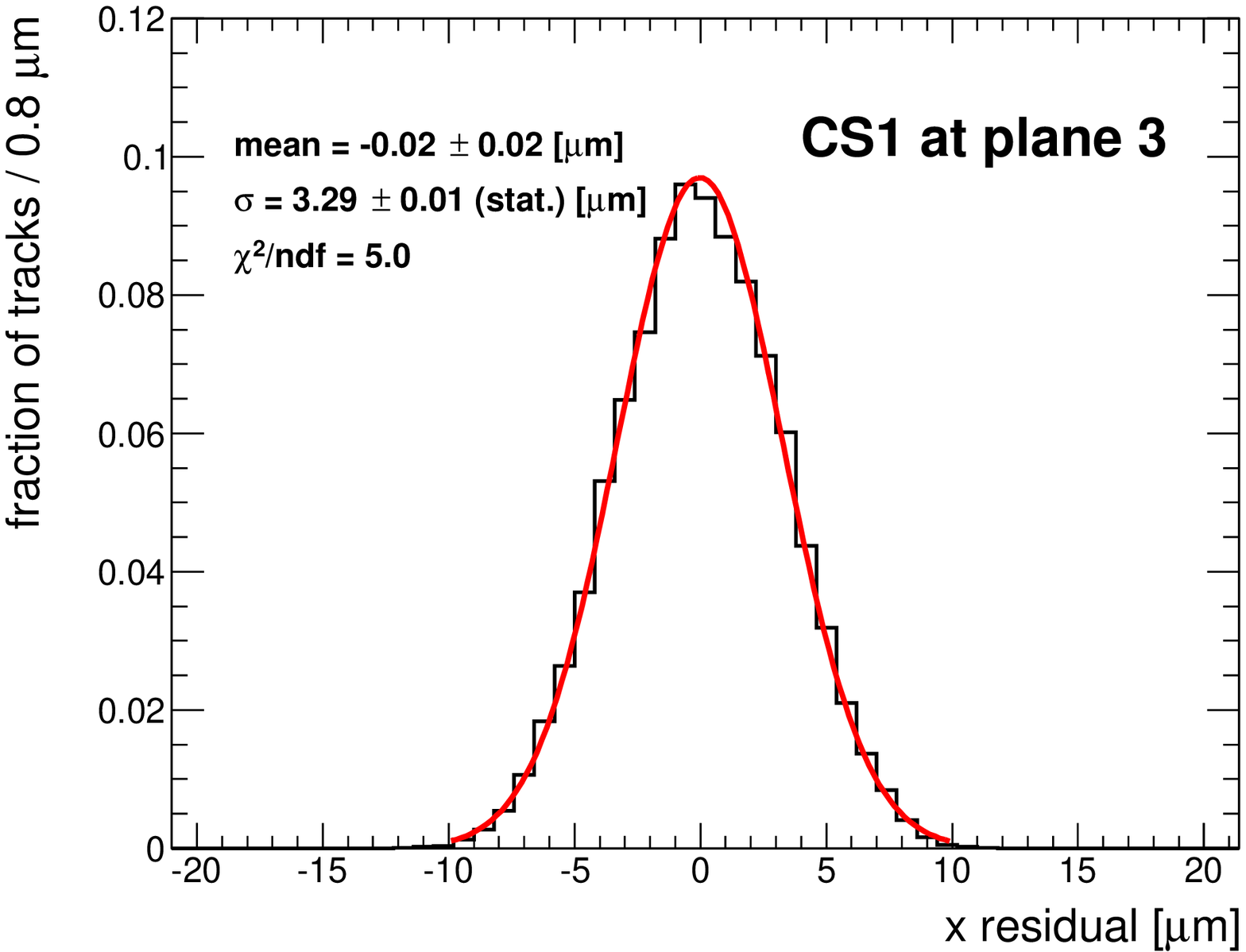}\put(-155,35){(a)}
	\includegraphics[width=.49\textwidth]{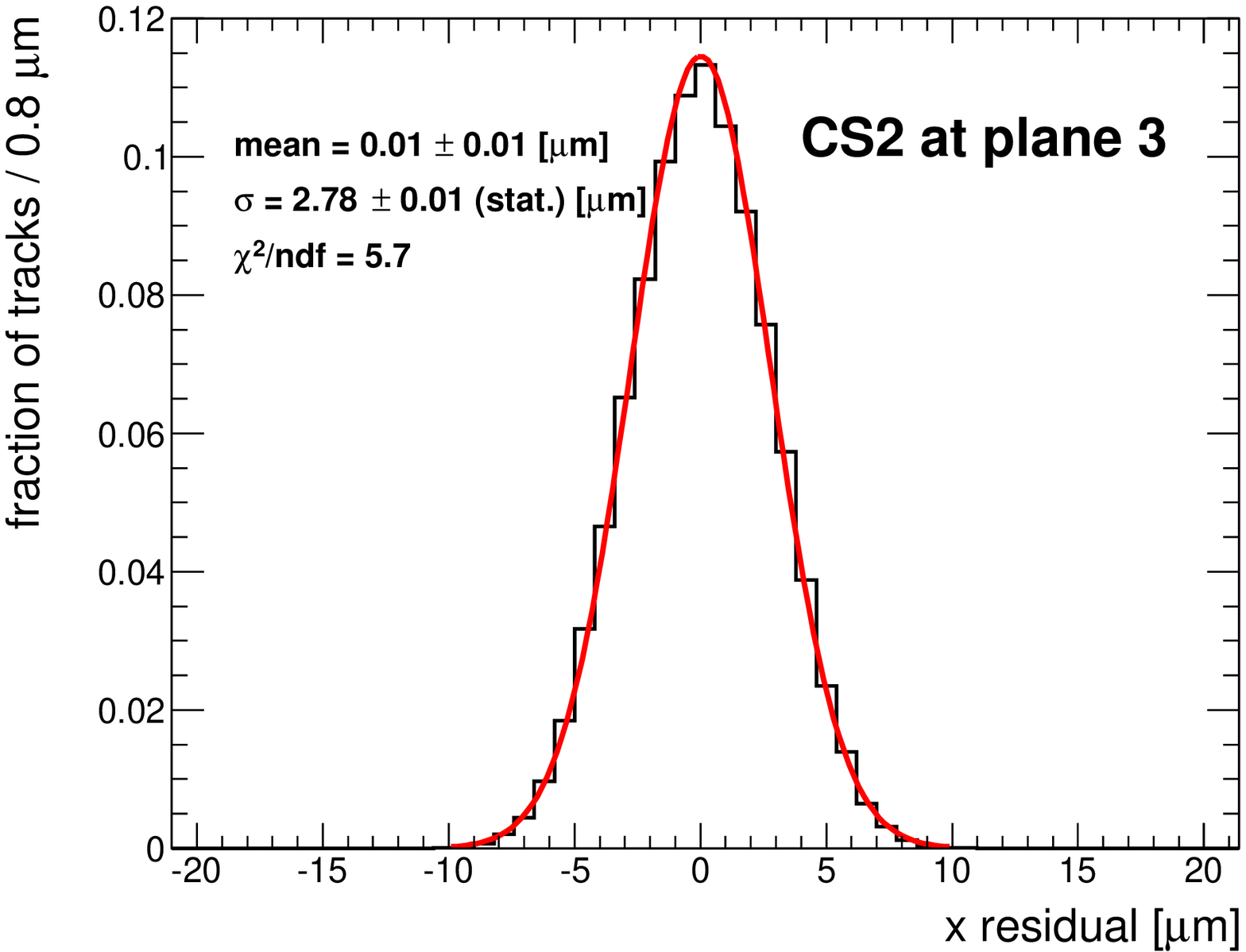}\put(-155,35){(b)}\\
	\includegraphics[width=.49\textwidth]{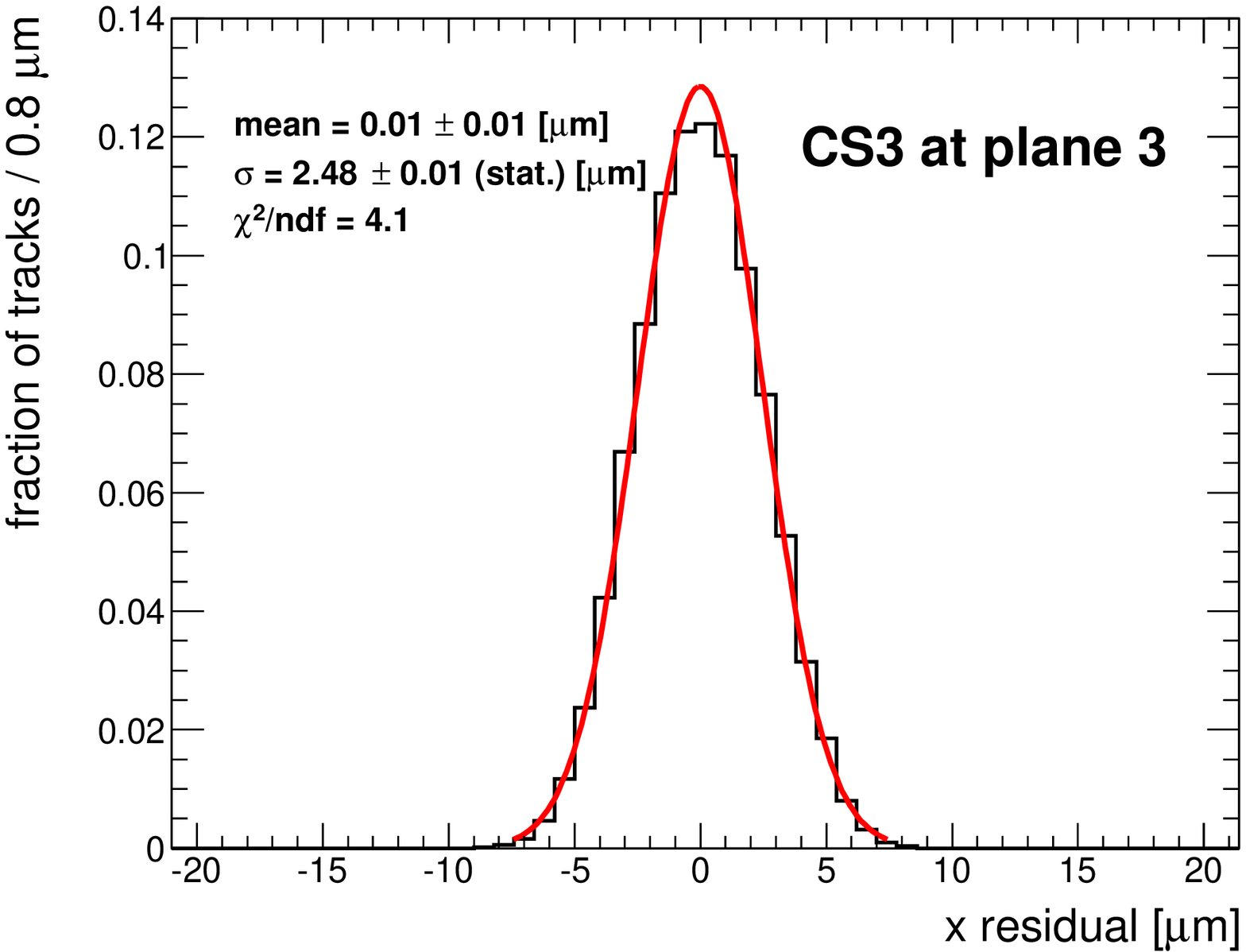}\put(-155,35){(c)}
	\includegraphics[width=.49\textwidth]{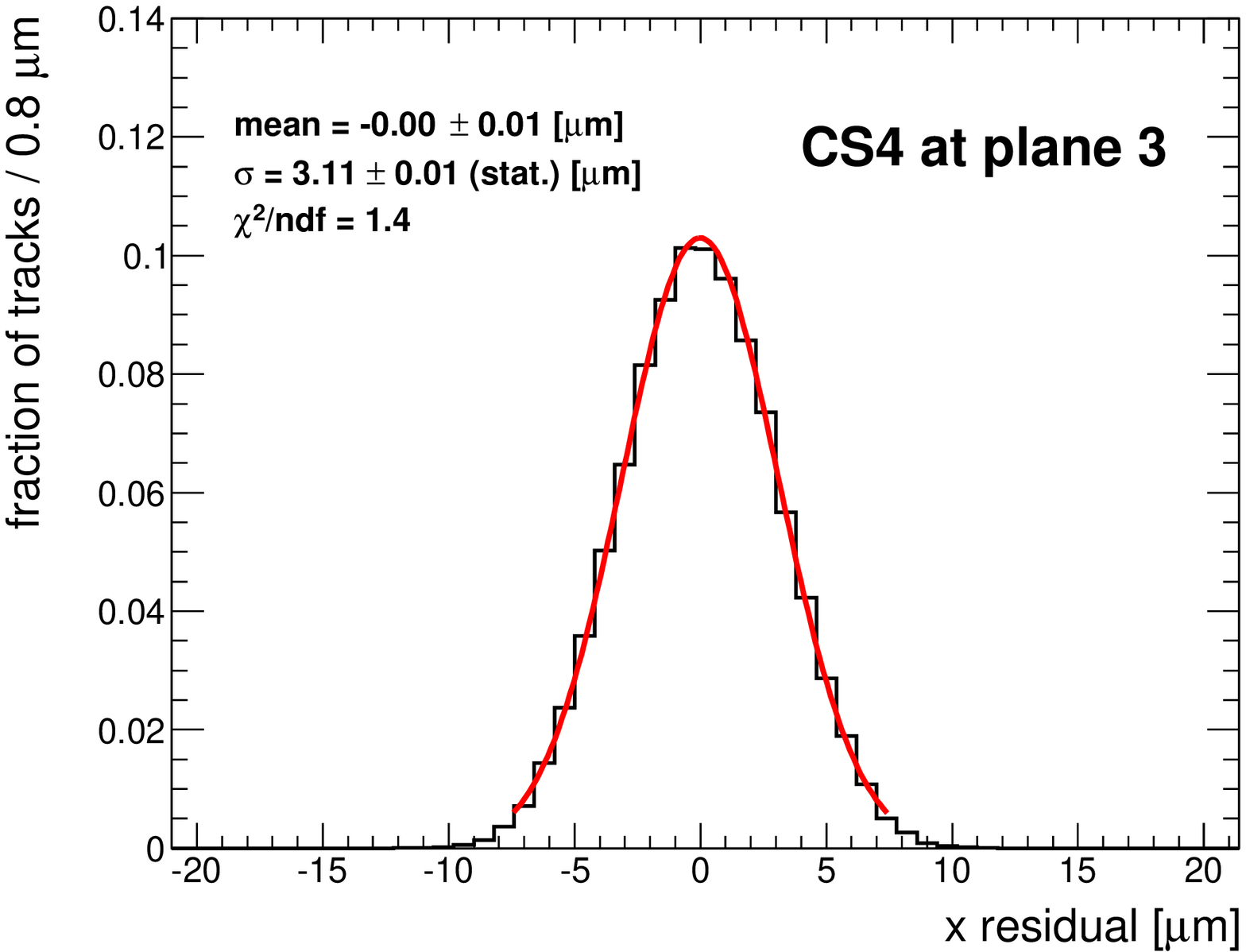}\put(-155,35){(d)}
	\else
	\includegraphics[width=.8\textwidth]{resi_CS1.eps}
	\includegraphics[width=.8\textwidth]{resi_CS2.eps}\\
	\includegraphics[width=.8\textwidth]{resi_CS3.eps}
	\includegraphics[width=.8\textwidth]{resi_CS4.eps}
	\fi
	\caption[Residual distribution]{The \textit{biased} residual distribution at a threshold of $\noise = 6$ for cluster sizes 1 to 4 is shown in (a) to (d), respectively.}
	\label{fig:resi_dist}
\end{figure}

\paragraph{Differential intrinsic resolution}
Following the discussion from reference~\cite{JansenEPJ}, an iterative pull analysis has been performed differentiating between cluster sizes. 
For each cluster size a pull distribution with a width equal to one is thereby obtained, the resulting biased residuals are shown in figure~\ref{fig:resi_dist}. 
The selection on the cluster size has only been applied to the plane under investigation. 
An intrinsic resolution of $(3.60\,\pm\,0.10\,(\textrm{sys.}))\,\micro\meter$, $(3.16\,\pm\,0.09\,(\textrm{sys.}))\,\micro\meter$, $(2.86\,\pm\,0.08\,(\textrm{sys.}))\,\micro\meter$,
 $(3.40\,\pm\,0.09\,(\textrm{sys.}))\,\micro\meter$ for cluster sizes 1 to 4, respectively, is found. 
The procedure of estimating the systematic uncertainties is discussed in reference~\cite{JansenEPJ}. 
The present geometry of the set-up combined with the given intrinsic resolution, a biased (unbiased) track resolution on plane\,3 of about $1.5\,\micro\meter$ ($1.7\,\micro\meter$) is derived \cite{gbltool},
 which allows for detailed intra-pixel studies.

\paragraph{Intra-pixel density distribution}

\begin{figure}[t]
	\center
	\ifdefined\notFOREPJ
	\includegraphics[width=.49\textwidth]{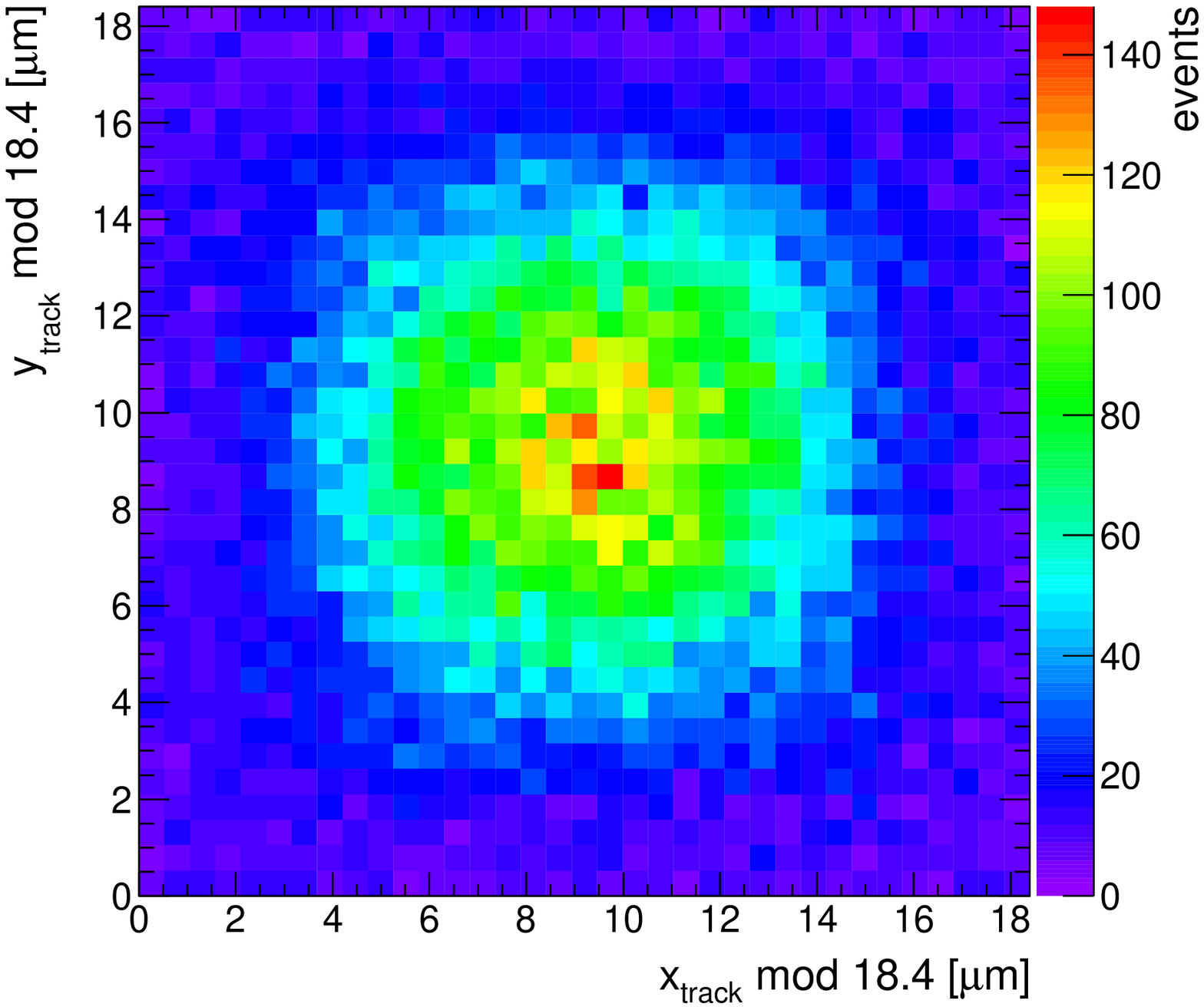}\put(-175,165){(a)}
	\includegraphics[width=.49\textwidth]{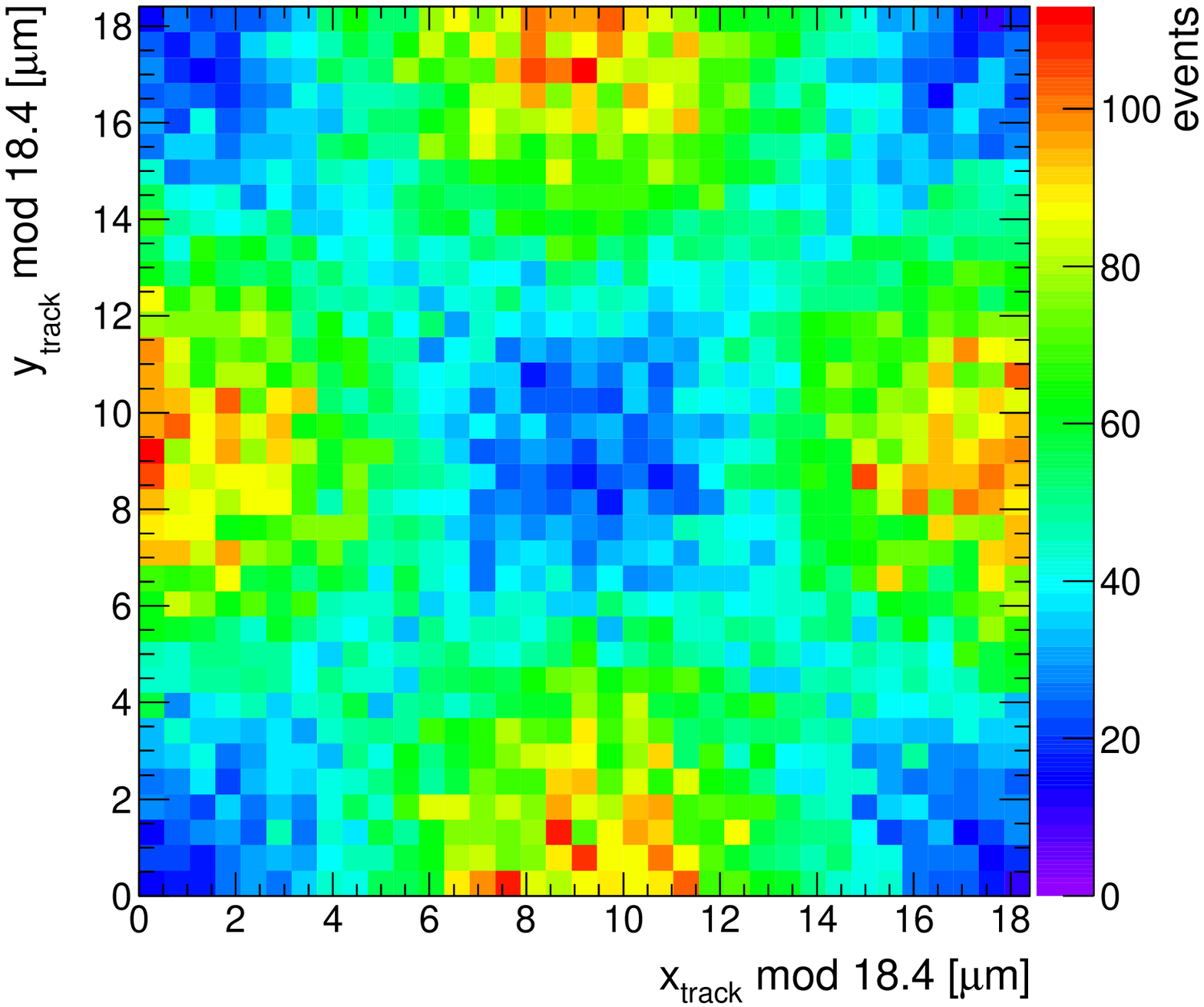}\put(-175,165){(b)}\\
	\includegraphics[width=.49\textwidth]{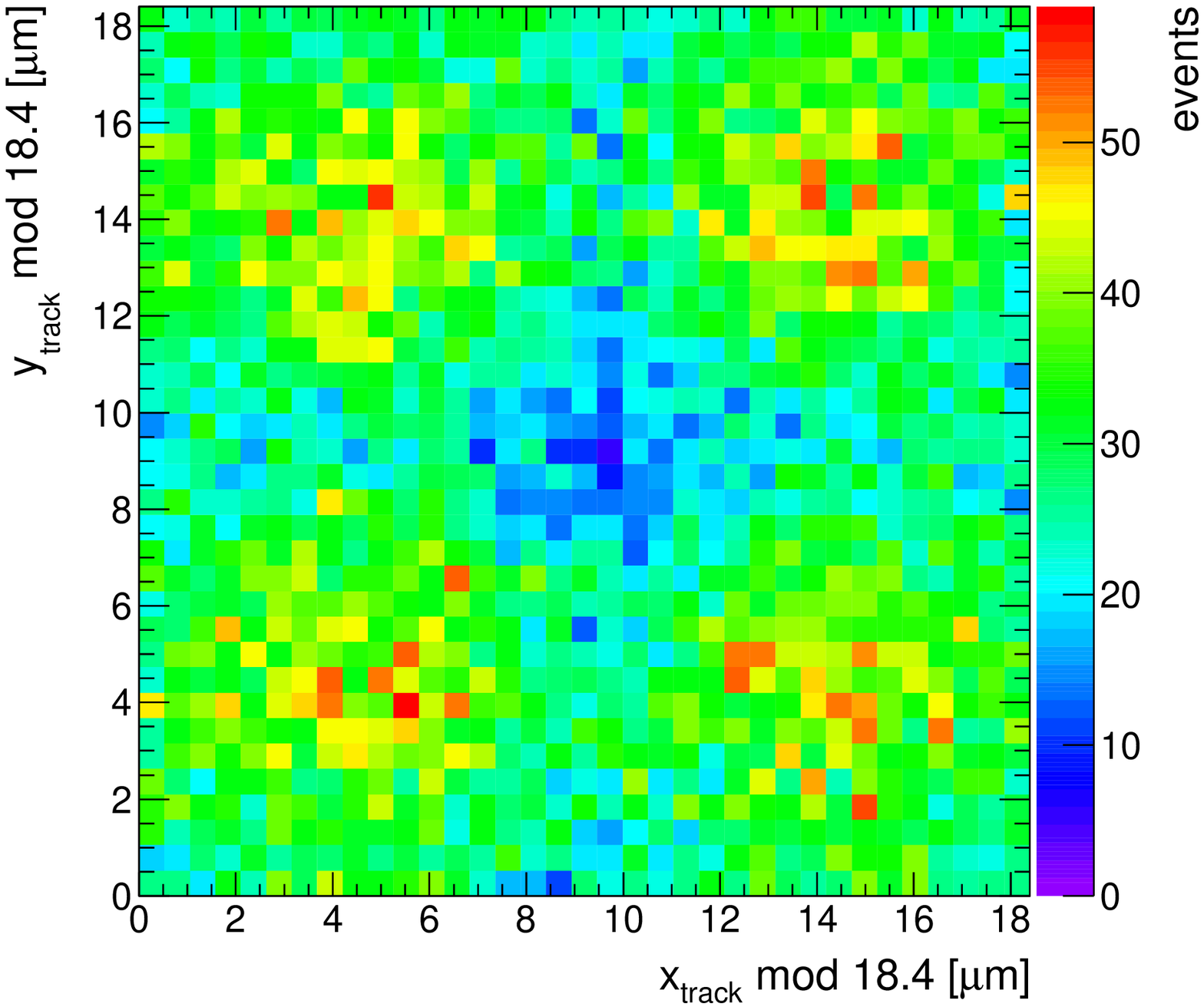}\put(-175,165){(c)}
	\includegraphics[width=.49\textwidth]{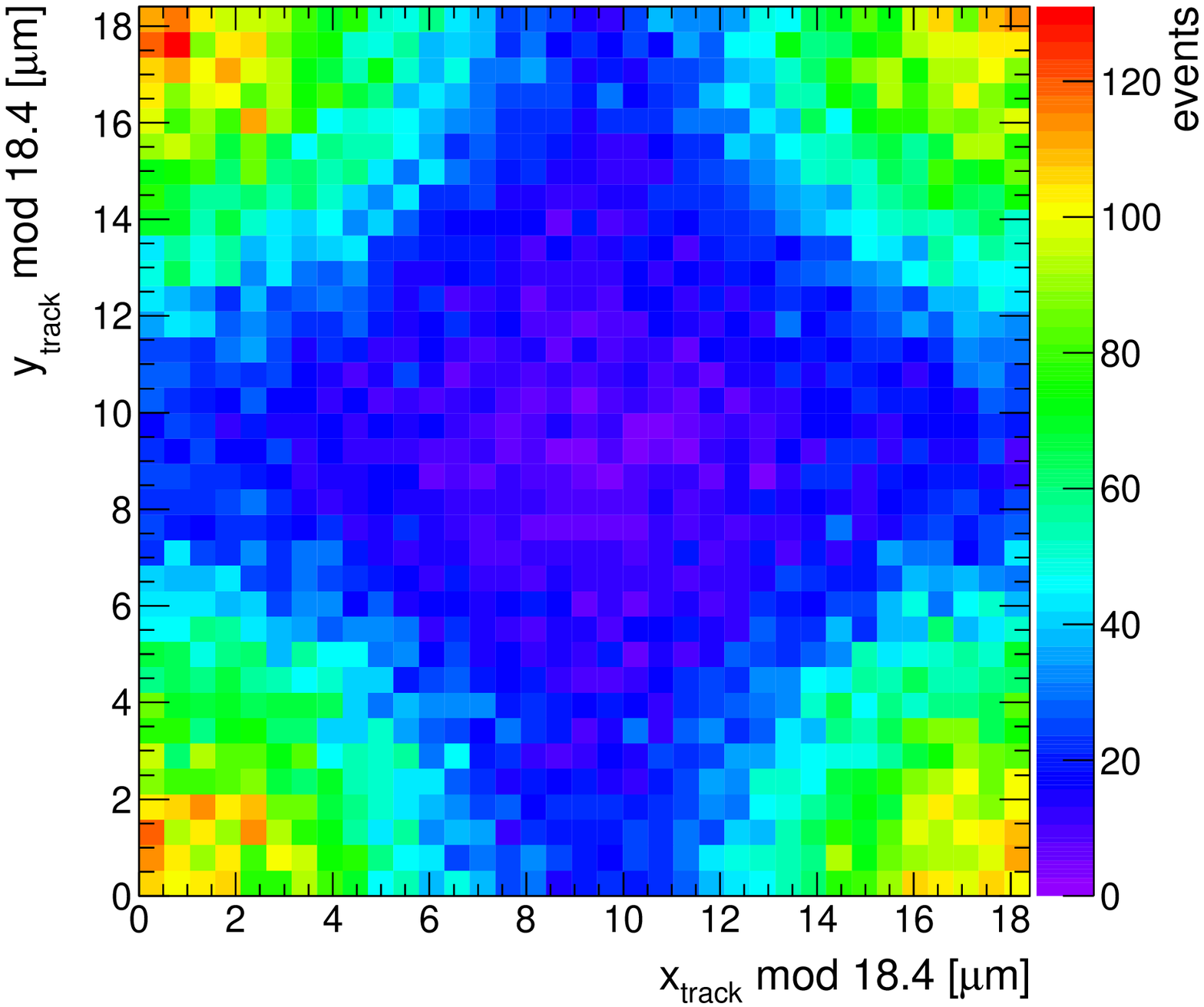}\put(-175,165){(d)}
	\else
	\includegraphics[width=.8\textwidth]{freq_CS1.eps}
	\includegraphics[width=.8\textwidth]{freq_CS2.eps}\\
	\includegraphics[width=.8\textwidth]{freq_CS3.eps}
	\includegraphics[width=.8\textwidth]{freq_CS4.eps}
	\fi
	\caption[intra-pixel density distribution]{The \textit{unbiased} intra-pixel density distribution at plane\,3 for cluster sizes 1 to 4 is shown in (a) to (d), respectively.}
	\label{fig:freq_dist}
\end{figure}

Figure~\ref{fig:freq_dist} shows the density of the \textit{unbiased}, reconstructed incident position as a function of the $x$- and $y$-coordinate on plane\,3 within a single pixel cell. 
In order to acquire sufficient statistics, tracks from the entire sensor have been mapped into a single pixel. 
This is done within the local coordinate system of plane\,3 performing a \textit{modulo} operation on the track incident position by multiples of the pixel size along the $x$- and $y$-axis. 
The origin of the local coordinate system coincides with the lower left corner of a physical pixel. 
For CS1, the distribution peaks in the centre of the pixel and falls off towards all sides. 
CS2-hits populate most likely the edges of the pixel cell, whereas CS4-hits usually result from tracks with an incident position close to a pixel corner. 
The density of CS3-hits peaks on the diagonal through the pixel cell at about $4\,\micro\meter$ along the $x$- and $y$-direction in the lower left quadrant, and accordingly in the others. 
Hence, the position of the peak density within a pixel can clearly be separated between the different cluster sizes. 

\paragraph{Intra-pixel residuals}
With a track resolution considerably smaller than the pixel size the construction of meaningful track residuals as a function of the track incident position are rendered possible. 
Figure~\ref{fig:sigxvsx_dist} depicts the residual-width as a function of the $x$-coordinate of \textit{biased} tracks. 
CS1-hits origin from incidents close to the pixel centre, as shown in figure~\ref{fig:freq_dist}\,(a). 
Since for CS1-hits the assumed position during tracking is the pixel centre, the residual-width is expected to have a minimum at the centre of the pixel,
 which is confirmed in figure~\ref{fig:sigxvsx_dist}\,(a).
Accordingly, CS2-hits have their minima at the edges of the pixel -- for clusters of size 2 along $x$ -- and in the middle -- for clusters of size 1 along $x$. 
For cluster sizes 3 and 4, larger uncertainties towards the centre of the pixel are observed, which results from the fact that only few tracks, that point to the centre of the pixel,
 result in hits with cluster size larger two. 
A discrepancy is observed in the position of the minimum for CS3:
The unbiased peak position from figure~\ref{fig:freq_dist}\,(c) is at about $4\,\micro\meter$ along the $x$-coordinate, but figure~\ref{fig:sigxvsx_dist}\,(c) shows a minimum at about $3\,\micro\meter$.
The assumed position of the CS3-hit within a pixel cell needs therefore to be corrected for this offset during tracking.

\begin{figure}[tb]
	\center
	\ifdefined\notFOREPJ
	\includegraphics[width=.49\textwidth]{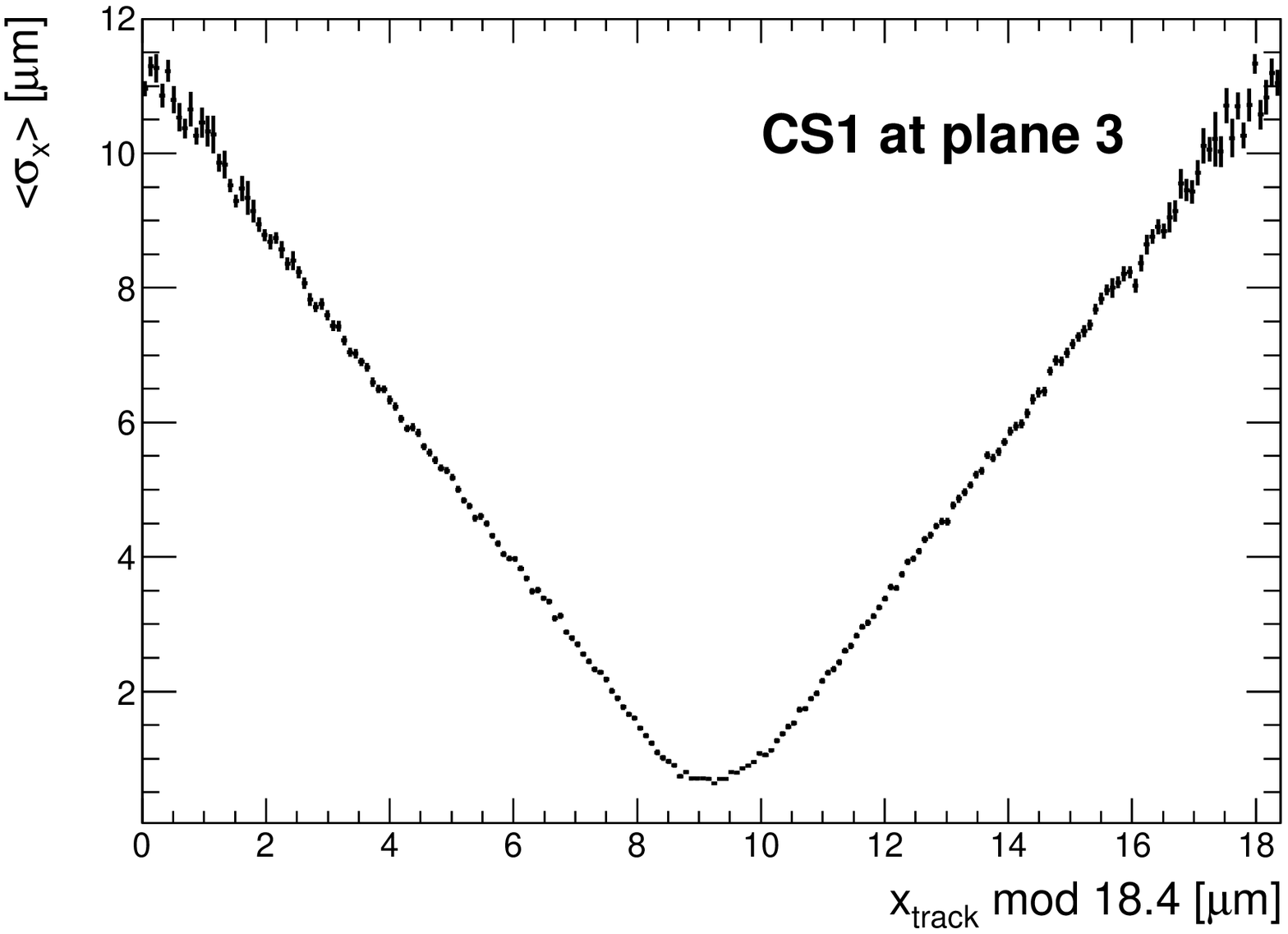}\put(-145,105){(a)}
	\includegraphics[width=.49\textwidth]{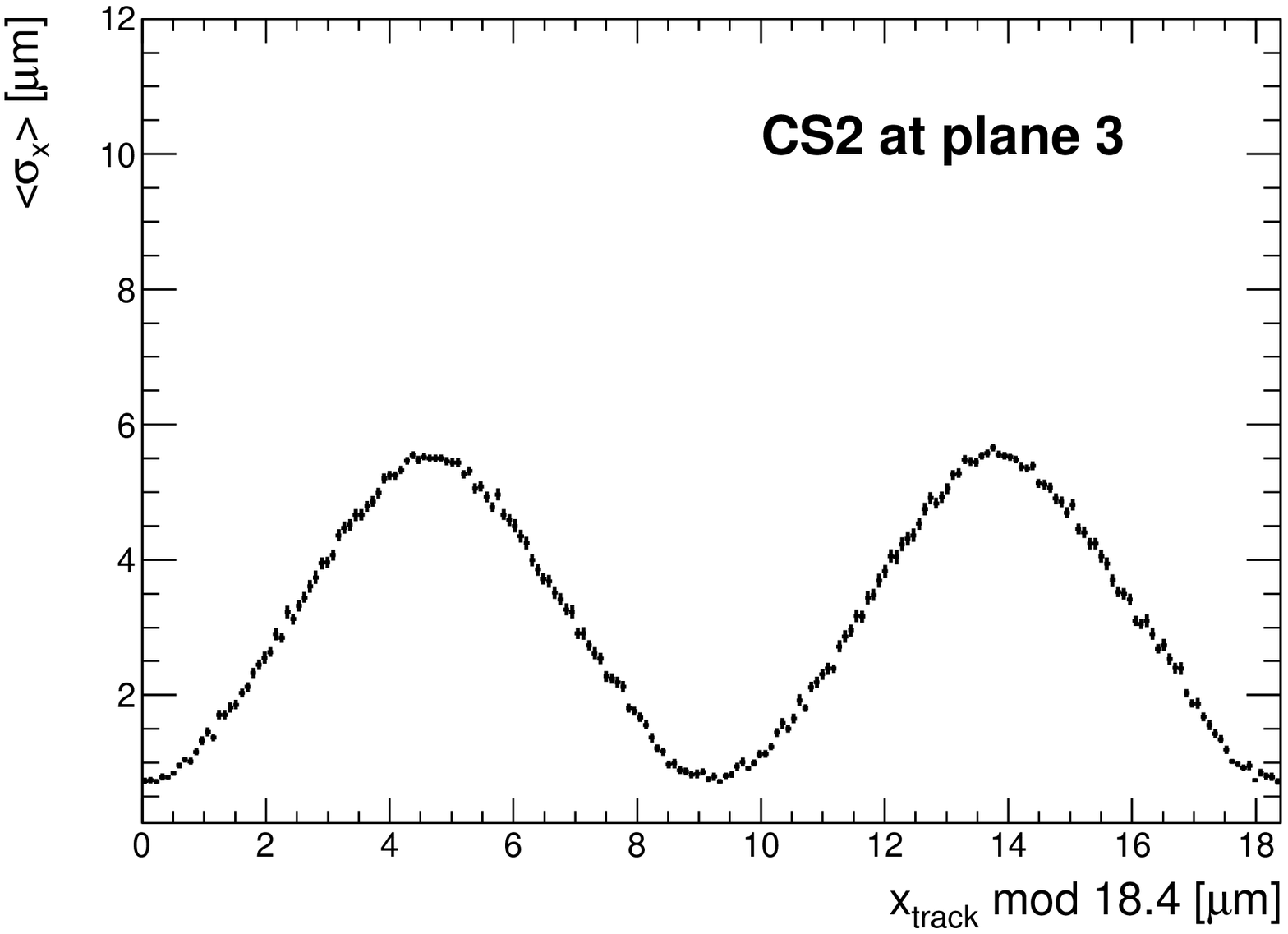}\put(-145,105){(b)}\\
	\includegraphics[width=.49\textwidth]{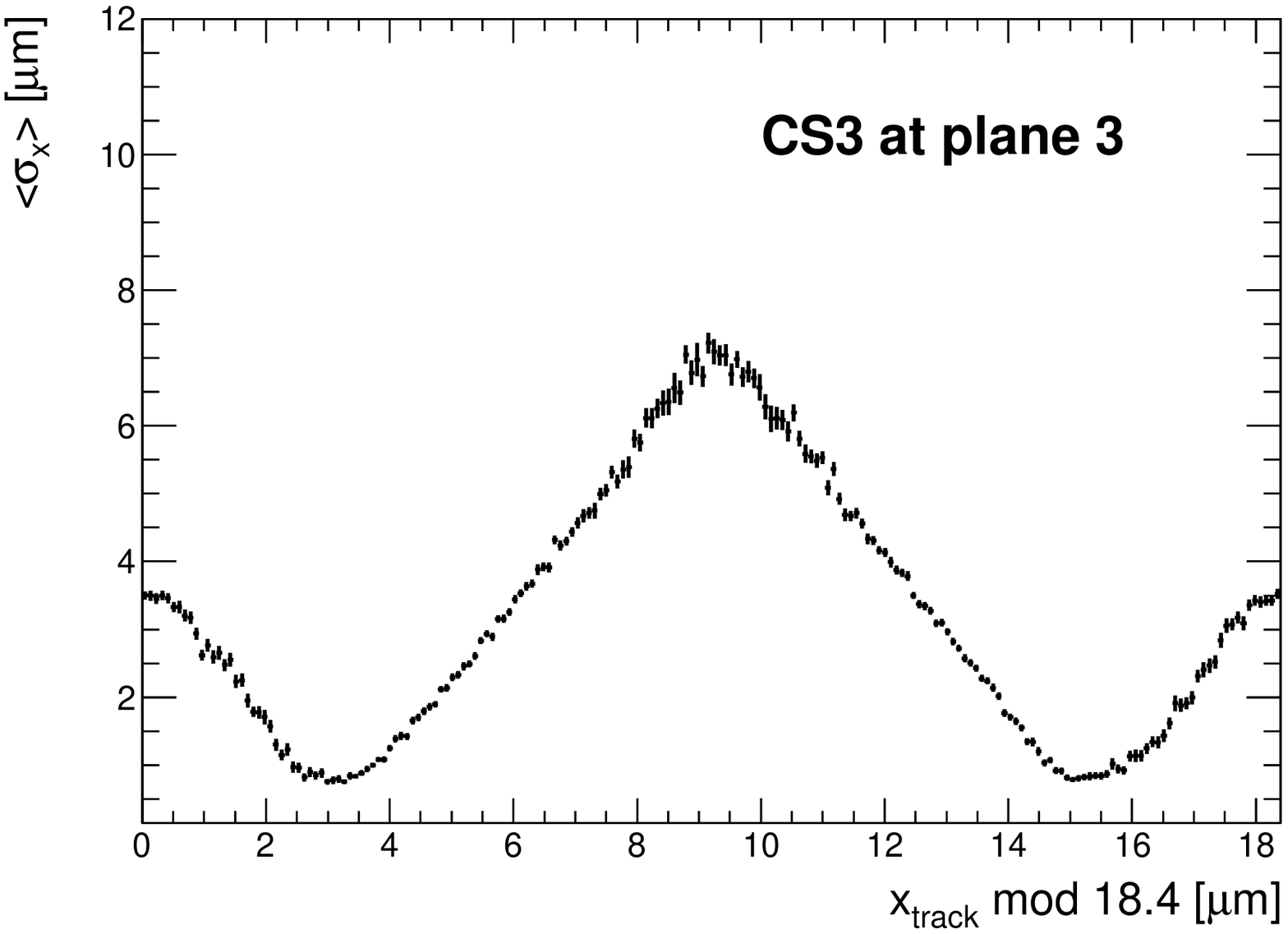}\put(-145,105){(c)}
	\includegraphics[width=.49\textwidth]{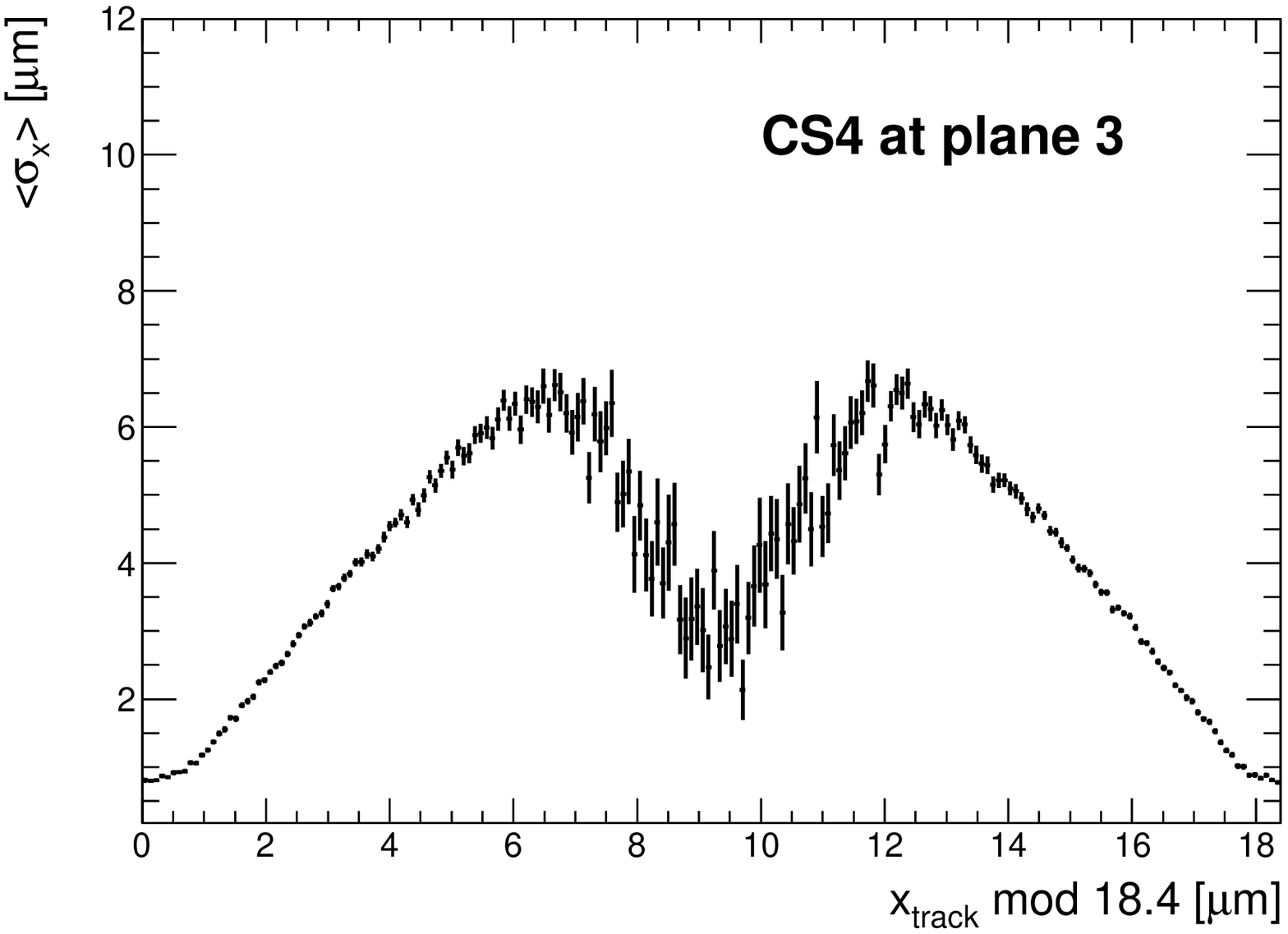}\put(-145,105){(d)}
	\else
	\includegraphics[width=.8\textwidth]{sigxvsx_CS1.eps}
	\includegraphics[width=.8\textwidth]{sigxvsx_CS2.eps}\\
	\includegraphics[width=.8\textwidth]{sigxvsx_CS3.eps}
	\includegraphics[width=.8\textwidth]{sigxvsx_CS4.eps}
	\fi
	\caption[intra-pixel residual distribution]{The \textit{biased} intra-pixel residual-width distribution at plane\,3 for cluster sizes 1 to 4 is shown in (a) to (d), respectively.}
	\label{fig:sigxvsx_dist}
\end{figure}

\else
 
\fi

\section{Conclusion}
\label{sec:conclusion}
\ifdefined\notFOREPJ
 
An analysis of test-beam data taken with the $\Datura$ beam telescope was presented discriminating between different sizes of clusters. 
As clusters produced by traversing particles differ from one another in terms of size and shape, also the accuracy of the reconstructed incident position depends on the original cluster. 
Investigating the different clustersizes individually allows for the extraction of the differential intrinsic resolution. 
With a biased track resolution of about $\unit{1.5}{\upmu\meter}$ on the telescope plane, intra-pixel studies revealed clearly distinguishable peak positions in density for different clustersizes
 and allowed for the analysis of the residual-width as a function of the track incident position within a pixel cell.

\else
 
\fi

\section*{Acknowledgements}
The test-beam support at DESY is highly appreciated. 
This work is supported by the Commission of the European Communities under the 6th Framework Programme 'structuring the European research area', contract number RII3-026126.
Furthermore, strong support from the Helmholtz Association and the BMBF is acknowledged.

{\small
\section*{Data and materials}
The datasets supporting the conclusions of this article are available from reference \cite{jansen_data}.
The software used is available from the github repositories: 1) \url{https://github.com/eutelescope/eutelescope}, 2) \url{https://github.com/simonspa/eutelescope/}, branch \textit{scattering}
 and 3) \url{https://github.com/simonspa/resolution-simulator}.


\bibliographystyle{IEEEtran}
\ifdefined\notFOREPJ
\bibliography{bibtex/refs}
\else
\bibliography{refs}

\begin{thebibliography}{10}
\providecommand{\url}[1]{#1}
\csname url@samestyle\endcsname
\providecommand{\newblock}{\relax}
\providecommand{\bibinfo}[2]{#2}
\providecommand{\BIBentrySTDinterwordspacing}{\spaceskip=0pt\relax}
\providecommand{\BIBentryALTinterwordstretchfactor}{4}
\providecommand{\BIBentryALTinterwordspacing}{\spaceskip=\fontdimen2\font plus
\BIBentryALTinterwordstretchfactor\fontdimen3\font minus
  \fontdimen4\font\relax}
\providecommand{\BIBforeignlanguage}[2]{{%
\expandafter\ifx\csname l@#1\endcsname\relax
\typeout{** WARNING: IEEEtran.bst: No hyphenation pattern has been}%
\typeout{** loaded for the language `#1'. Using the pattern for}%
\typeout{** the default language instead.}%
\else
\language=\csname l@#1\endcsname
\fi
#2}}
\providecommand{\BIBdecl}{\relax}
\BIBdecl

\bibitem{EUDETwp}
N.~Potylitsina-Kube \emph{et~al.}, ``{EUDET},'' \url{http://www.eudet.org},
  accessed: 26.07.2016.

\bibitem{ref:eudetreport200902}
A.~Bulgheroni, ``Results from the {EUDET} telescope with high resolution
  planes,'' \url{https://www.eudet.org/e26/e27/e50990/eudet-report-09-02.pdf},
  Tech. Rep. EUDET-Report-2009-02, 2009, accessed: 21.04.2015.

\bibitem{JansenEPJ}
\BIBentryALTinterwordspacing
H.~Jansen \emph{et~al.}, ``{{Performance of the EUDET-type beam telescopes}},''
  \emph{EPJ Techn. Instrum.}, vol.~3, no.~1, p.~7, 2016, {{DESY-16-055,
  arXiv:1603.09669}}. [Online]. Available:
  \url{10.1140/epjti/s40485-016-0033-2}
\BIBentrySTDinterwordspacing

\bibitem{Spannagel201671}
\BIBentryALTinterwordspacing
S.~Spannagel, ``{Status of the CMS Phase I pixel detector upgrade},''
  \emph{Nucl. Instr. Meth. Phys. Res. A}, vol. 831, pp. 71 -- 75, 2016,
  proceedings of the 10th International “Hiroshima” Symposium on the
  Development and Application of Semiconductor Tracking Detectors. [Online].
  Available: \url{10.1016/j.nima.2016.03.028}
\BIBentrySTDinterwordspacing

\bibitem{AlipourTehrani:2133128}
\BIBentryALTinterwordspacing
A.~Tehrani \emph{et~al.}, ``{Test beam analysis of ultra-thin hybrid pixel
  detector assemblies with Timepix readout ASICs},'' Feb 2016,
  {CLICdp-Note-2016-001}. [Online]. Available:
  \url{https://cds.cern.ch/record/2133128}
\BIBentrySTDinterwordspacing

\bibitem{1748-0221-10-03-C03044}
\BIBentryALTinterwordspacing
H.~Augustin \emph{et~al.}, ``{The MuPix high voltage monolithic active pixel
  sensor for the Mu3e experiment},'' \emph{JINST}, vol.~10, no.~03, p. C03044,
  2015. [Online]. Available: \url{10.1088/1748-0221/10/03/C03044}
\BIBentrySTDinterwordspacing

\bibitem{bib:IBLprototypes}
\BIBentryALTinterwordspacing
{ATLAS IBL Collaboration}, ``{Prototype ATLAS IBL Modules using the FE-I4A
  Front-End Readout Chip},'' \emph{JINST}, vol.~7, p. P11010, 2012. [Online].
  Available: \url{10.1088/1748-0221/7/11/P11010}
\BIBentrySTDinterwordspacing

\bibitem{bib:AFP3D2}
\BIBentryALTinterwordspacing
J.~Lange, E.~Cavallaro, S.~Grinstein, and I.~Lopez~Paz, ``{3D silicon pixel
  detectors for the ATLAS Forward Physics experiment},'' \emph{JINST}, vol.~10,
  p. C03031, 2015. [Online]. Available: \url{10.1088/1748-0221/10/03/C03031}
\BIBentrySTDinterwordspacing

\bibitem{EUDET-2007-11}
T.~Behnke \emph{et~al.}, ``{Test Beams at DESY},''
  \url{http://www.eudet.org/e26/e28/e182/e283/eudet-memo-2007-11.pdf}, Tech.
  Rep., 2007, accessed: 21.04.2015.

\bibitem{HuGuo2010480}
\BIBentryALTinterwordspacing
C.~Hu-Guo \emph{et~al.}, ``First reticule size {MAPS} with digital output and
  integrated zero suppression for the {EUDET-JRA1} beam telescope,''
  \emph{Nucl. Instrum. Methods Phys. Rev. A}, vol. 623, no.~1, pp. 480 -- 482,
  2010, 1st International Conference on Technology and Instrumentation in
  Particle Physics. [Online]. Available: \url{10.1016/j.nima.2010.03.043}
\BIBentrySTDinterwordspacing

\bibitem{Spartan3}
{XILINX INC.}, ``{Spartan-3E Starter Kit},''
  \url{https://www.xilinx.com/products/boards-and-kits/hw-spar3e-sk-us-g.html},
  accessed: 04.11.2016.

\bibitem{Blobel20111760}
\BIBentryALTinterwordspacing
V.~Blobel, C.~Kleinwort, and F.~Meier, ``Fast alignment of a complex tracking
  detector using advanced track models,'' \emph{Computer Physics
  Communications}, vol. 182, no.~9, pp. 1760 -- 1763, 2011, computer Physics
  Communications Special Edition for Conference on Computational Physics
  Trondheim, Norway, June 23-26, 2010. [Online]. Available:
  \url{10.1016/j.cpc.2011.03.017}
\BIBentrySTDinterwordspacing

\bibitem{Kleinwort-2012}
\BIBentryALTinterwordspacing
C.~{Kleinwort}, ``{General broken lines as advanced track fitting method},''
  \emph{Nucl. Instr. Meth. Phys. Res. A}, vol. 673, pp. 107--110, May 2012.
  [Online]. Available: \url{10.1016/j.nima.2012.01.024}
\BIBentrySTDinterwordspacing

\bibitem{gbltool}
\BIBentryALTinterwordspacing
S.~Spannagel and H.~Jansen, ``{GBL Track Resolution Calculator},'' accessed:
  03.03.2016. [Online]. Available:
  \url{https://github.com/simonspa/resolution-simulator}
\BIBentrySTDinterwordspacing

\bibitem{jansen_data}
H.~Jansen, ``{Dataset for the 'Performance of the EUDET-type beam telescopes'
  publication (EPJ TI)},'' \url{10.5281/zenodo.59255}, Aug. 2016.

\end{thebibliography}
\fi
}

\end{document}